\documentclass[a4,12pt]{article}
\usepackage{epsfig}
\usepackage{amssymb}

\newcommand{\be}{\begin{equation}}
\newcommand{\ee}{\end{equation}}
\newcommand{\bea}{\begin{eqnarray}}
\newcommand{\eea}{\end{eqnarray}}

\begin{document}
\title{Continuum limit of the spectrum of the hadronic string}
\author{Pushan Majumdar\thanks{e-mail:pushan.majumdar@uni-graz.at}\\
{\em Institut f\"ur Physik, Fachbereich Theoretische Physik,} \\
{\em Karl Franzens Universit\"at, Graz, \"Osterreich}}
\maketitle
\begin{abstract}
We look at the continuum limit of the two universal predictions of the hadronic 
string theory namely the L\"uscher term and the energy difference between successive
energy states in three dimensional SU(2) lattice gauge theory using Wilson loops
and Polyakov loop correlators. We also compare our simulation data with the theoretical
predictions in \cite{LW2}.
\end{abstract}

\section{Introduction}

Formation of a flux tube between quarks and anti-quarks in the QCD vacuum is an 
appealing mechanism for quark confinement. Effective theories for these flux tubes 
can be written in terms of string theories \cite{str1}. An interesting feature of
these theories is that they predict a universal coefficient of the large distance
$1/r$ term $c$ ($=-\frac{\pi}{24}(d-2); \;d:$ no. of space-time dimensions) 
\cite{c} 
also known as the L\"uscher term. While ground states of such string theories have 
been studied for a long time \cite{strold}, now with improvement of both computing 
power and algorithms \cite{LW}, it is possible to do careful studies of other 
important properties of the flux tube and several 
such studies have been carried out recently \cite{str2}. 

In a previous work we had applied the L\"uscher-Weisz multilevel algorithm
to look at the spectrum of the hadronic string \cite{P1}. Since that algorithm 
allows an exponential error reduction, we were able to look at physically large 
time extents for the Wilson loops. In this paper we apply the same algorithm to
 study the 
continuum limit of two of the string observables, the universal coefficient ``$c$"
and the energy difference between the ground state and the 
first excited state. Our previous experience had hinted that systematic 
corrections to the energies of the string states arising from the presence of higher excited states
played a very crucial role in the determination of the energy differences between 
successive states. 
Recently there have been extensive studies of the string spectrum using asymmetric lattices 
and sophisticated wave functions \cite{Kuti}. Such studies have suggested a hyperfine structure
of the spectrum. Here we use completely independent and in some sense complimentary methods to
look at the string spectrum.

\section{Simulation parameters}

We carried out simulations of three dimensional SU(2) lattice gauge theory on three
different lattices chosen so as to roughly cover the same physical volume. Let us
call these lattices {\bf A, B} and {\bf C}. We restrict ourselves to square lattices
 whose parameters are summarized
in table \ref{tab0}. The lattice spacing $a$ was obtained by assuming $\sqrt{\sigma}
=(0.5 \,{\rm fm})^{-1}$. On all these 
lattices, we computed Polyakov loop correlators 
$\langle P^*(x)P(y)\rangle$ for various values of $r=y-x$ and Wilson loops of 
various space and time extents $\langle W(r,T)\rangle$. 

\begin{center}
\begin{table}[htb]\caption{Lattice parameters}\label{tab0}
\begin{tabular}{cccllll}
\hline
$lattice$ & $\beta$ & $L/a$ & $a\sqrt{\sigma}$ & $r_0/a$ & $a$ (fm) & 
$\sqrt{\sigma} r_0$ \\
\hline
{\bf A} & 7.5 & 36 & 0.19622 (2) & 6.285 (4) & 0.09811 (1) & 1.2332 (10) \\
{\bf B} & 10.0 & 48 & 0.143691 (14) & 8.578 (4) & 0.071846 (7) & 1.2319 (11) \\
{\bf C} & 12.5 & 60 & 0.113243 (5) & 10.92 (3) & 0.056622 (3) & 1.2366 (34) \\
\hline
\end{tabular}
\end{table}
\end{center}

To reliably extract signals of these observables which are exponentially 
decreasing functions of $r$ and $T$, we used the L\"uscher-Weisz multilevel
algorithm \cite{LW}. As is well known, this algorithm has several optimization 
parameters
 among which the number of sub-lattice updates employed seem to be the most 
important one \cite{P2}. In tables \ref{t1} and \ref{t2} we tabulate the number of 
sub-lattice updates (iupd) used in our measurements.  

\begin{center}
\begin{table}[htb]\caption{Sublattice updates for Polyakov
loop correlators}\label{t1}
\begin{tabular}{c|ccc|cc|ccc}
\hline
$lattice$ &\multicolumn{3}{c|}{\bf A} &
\multicolumn{2}{c|}{\bf B} & \multicolumn{3}{c}{\bf C} \\
\hline
$r$ & 4$-$8 & 7$-$11 &10$-$14 & 6$-$12 & 11$-$16 & 8$-$13 & 11$-$16 & 15$-$20 \\
iupd & 2000 & 3200 & 8000 & 16000 & 32000 & 8000 & 16000 & 32000 \\
\hline
\end{tabular}
\end{table}
\end{center}

\begin{table}[htb]\caption{Sublattice updates for Wilson loops}\label{t2}
\begin{center}
\vspace*{-5mm}\begin{tabular}{ccc|ccc|ccc}
\hline
\multicolumn{3}{c|}{$lattice$ {\bf A}}&\multicolumn{3}{c|}{$lattice$ {\bf B}}&
\multicolumn{3}{c}{$lattice$ {\bf C}} \\
\hline
$T$ & $r$ & iupd & $T$ & $r$ & iupd & $T$ & $r$ & iupd \\
\hline
4 & 4$-$8 & 300 & 4 & 8$-$12 & 300 & 6 & 8$-$12 & 200 \\
  & 8$-$12 & 400 &  & 12$-$16 & 400 &  & 12$-$16 & 250 \\
  &        &     &  &         &     &  & 16$-$20 & 300 \\
&&&&&&&& \\
6 & 4$-$8 & 450 & 8 & 8$-$12 & 600 & 12 & 8$-$12 & 400 \\
  & 8$-$12& 600 &   & 12$-$16& 800 &    & 12$-$16& 500 \\
  &       &     &   &        &     &    & 16$-$20& 600 \\
&&&&&&&& \\
12& 4$-$8 & 900 & 12& 8$-$12 & 900 & 20 & 8$-$12 & 800 \\
  & 8$-$12& 1200&   & 12$-$16& 1200&    & 12$-$16& 1000 \\
  &       &     &   &        &     &    & 16$-$20& 1200 \\
&&&&&&&& \\
18& 4$-$8 & 1350& 16& 8$-$12 & 1200& 30 & 8$-$12 & 1000 \\
  & 8$-$12& 1800&   & 12$-$16& 1600&    & 12$-$16& 1250 \\
  &       &     &   &        &     &    & 16$-$20& 1500 \\
\hline
\end{tabular}
\end{center}
\end{table}

Another important parameter is the thickness of the time-slice over which the
sub-lattice averages are carried out. 
We found that it was helpful to increase this thickness as
one goes from stronger to weaker coupling.  
While a thickness of 2 was sufficient at 
$\beta=5.0$ \cite{P1}, for $\beta=10$ we chose the thickness to be 4 uniformly while 
for the 
other two $\beta$ values, 7.5 and 12.5, we chose the higher value between 4 and 6 whenever 
possible.

Only one level of averaging was sufficient for most of the observables. However
to get a stable value of the L\"uscher term it was necessary to do a second level
of averaging of the Polyakov loop data before taking the derivatives especially at 
higher values of $\beta$.

\section{Polyakov loop}

\begin{figure}
\vspace*{-2.0cm}
\begin{center}
\mbox{\epsfig{file=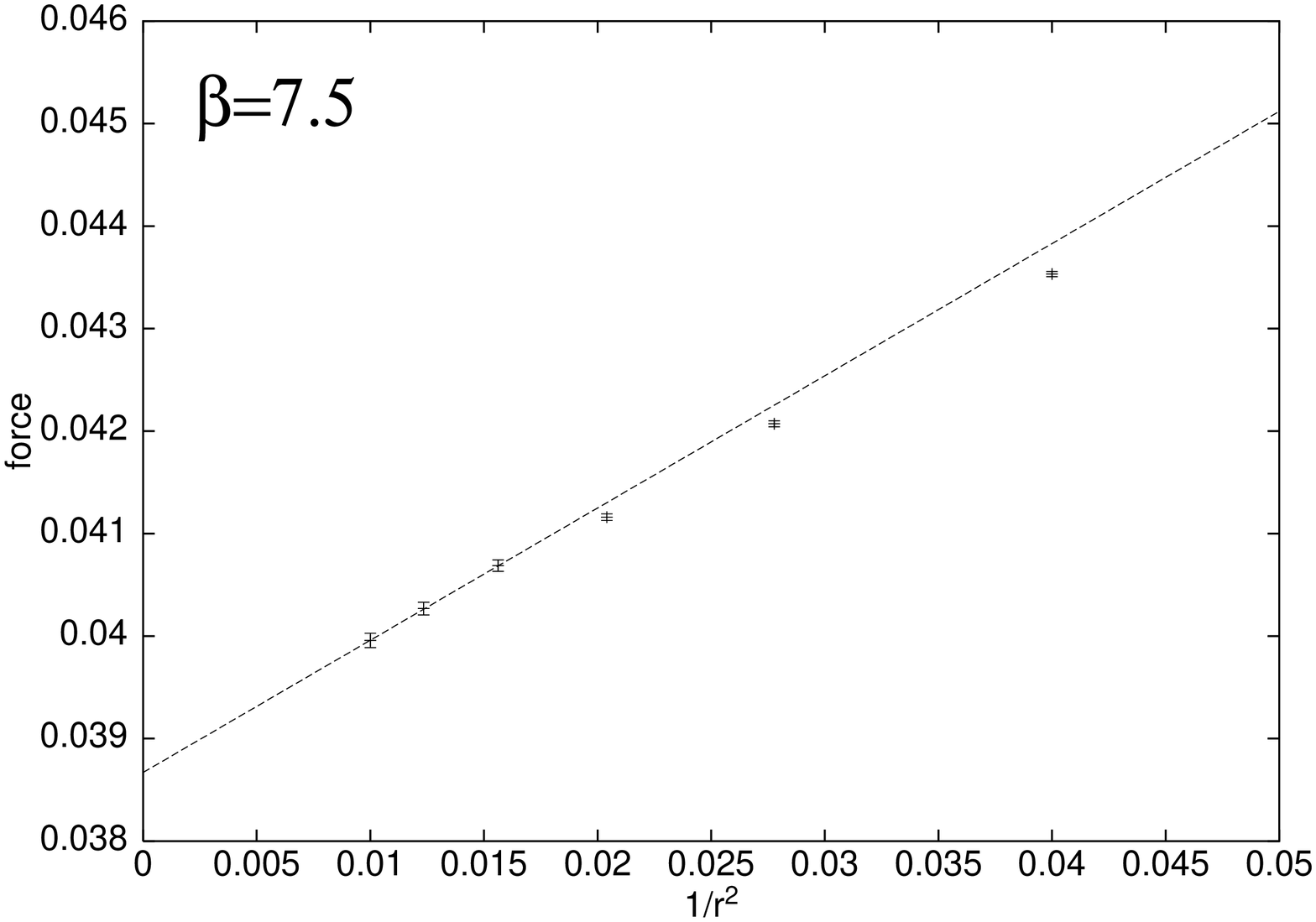,width=10truecm,angle=0}}
\mbox{\epsfig{file=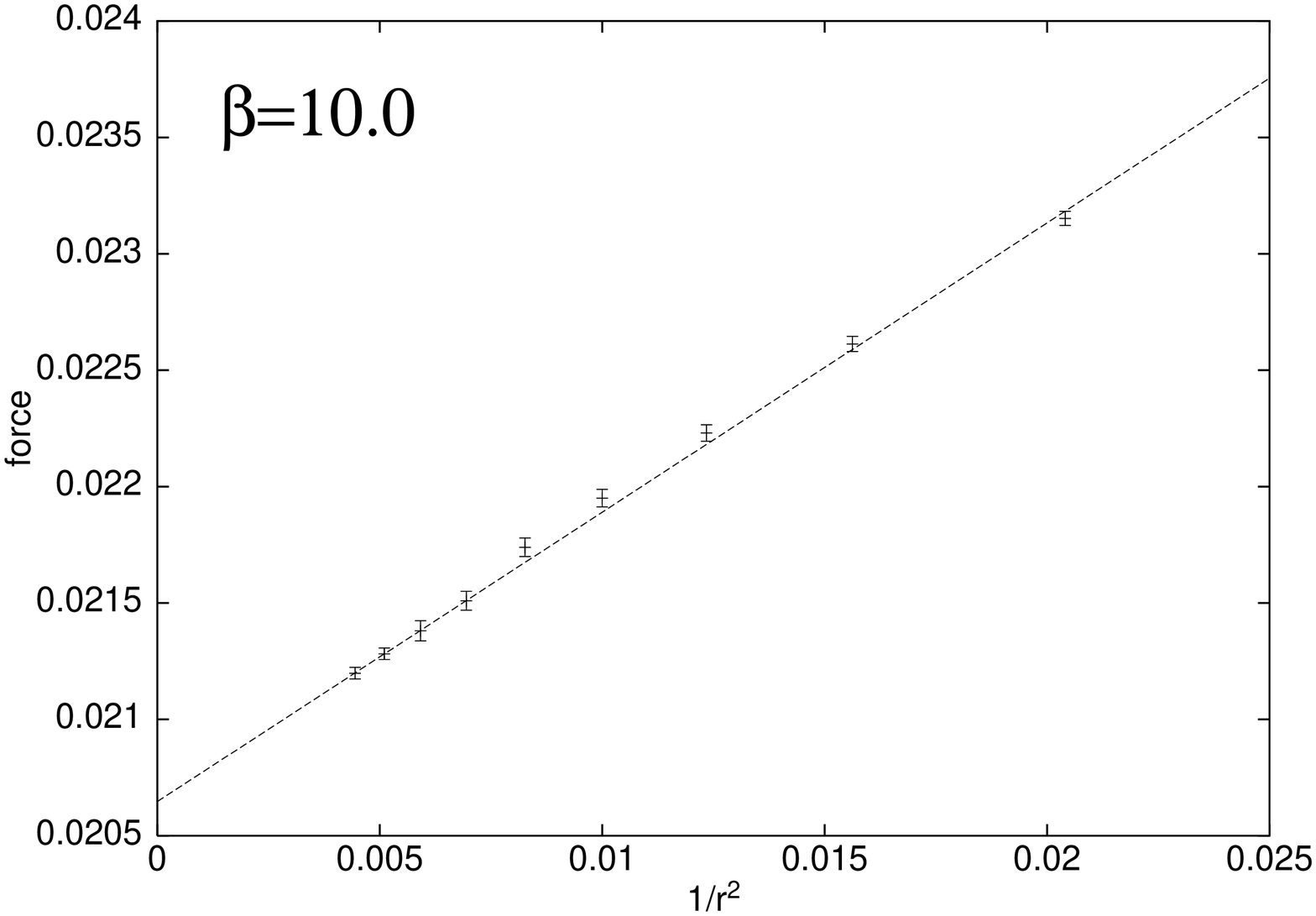,width=10truecm,angle=0}}
\mbox{\epsfig{file=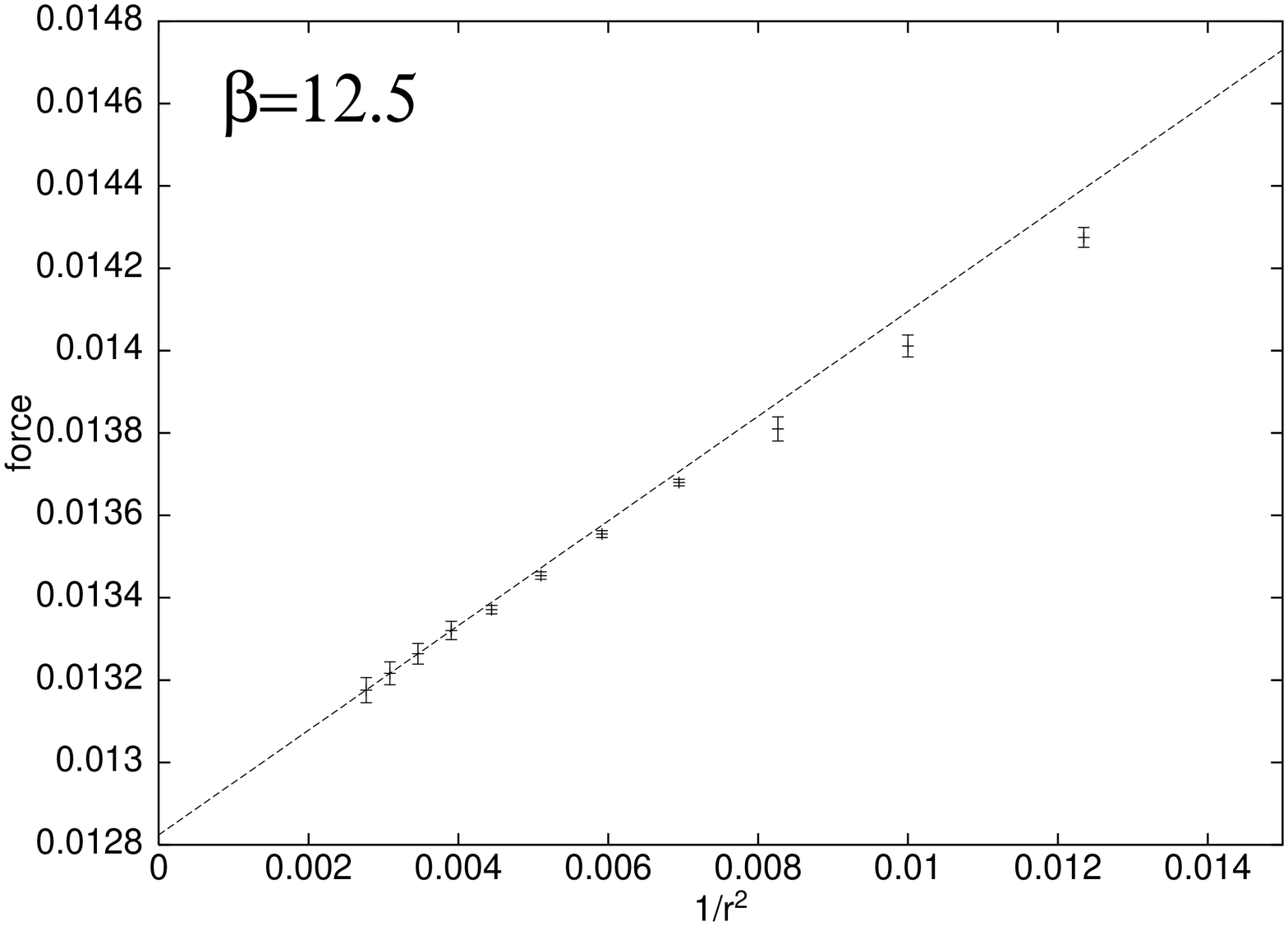,width=10truecm,angle=0}}
\caption{Forces at various values of $\beta$}\label{force.fig}
\end{center}
\end{figure}

The Polyakov loop correlator can be expressed as
\be
\langle P^*(r,T)P(0,T)\rangle = \sum_{i=0}^{\infty} b_i \exp[-V_i(r)T]
\ee
where $b_i$'s are integers and $T=L$ is the full time extent of the lattice.
For such large values of $T$ this expectation value projects almost exclusively
 to the ground state and we estimate that corrections due to higher states are lower
than our statistical errors. From the Polyakov loop correlators we determine
the potential $V(r)$ and the force $F(r)$ as
\bea
V(r)&=&-\frac{1}{T}\log\langle P^*(r,T)P(0,T)\rangle \\
F(r)&=& \frac{\partial V(r)}{\partial r} \equiv (V(r+1)-V(r-1))/2 .
\eea 
The string tension is determined from the intercept of the curve of $F(r)$
versus $1/r^2$. This will yield the correct value
only for asymptotically large distances as corrections $\propto r^{-3}$ become
negligible. However in our simulations at moderate values of $r$  these
corrections are not yet negligible. Thus for us $F(r)$ is not really a
straight line but one which approaches a straight line close to the origin.
To avoid the short distance effects in our determination of the string tension as
much as possible, we fit a straight line only to the final few points closest to
the origin and take the intercept from that fit. These fits are shown in figure
\ref{force.fig}.

To have a theoretical prediction against which we can compare the simulation 
results, we will consider two different forms of the potential. These are the 
potential assuming a free bosonic string description of the flux tube and the
Arvis potential \cite{Arvis} which comes from assuming a Nambu-Goto type description 
of the 
flux tube. While the free bosonic string is the simplest model, recent studies
\cite{LW2} suggest that the only consistent description of the Polyakov loop 
correlator is
given by the truncated Arvis potential. At large distances, predictions from both these models 
coincide. Our comparison will be at moderate distances to see which form is 
preferred.

For the free bosonic string in three dimensions, we have $V(r)$ to be of the form 
\be 
V(r) = \sigma r + {\hat V} - \pi/24r 
\ee
while the Arvis potential is given by 
\be
V(r) = \sigma r\left ( 1-\frac{\pi}{12 \sigma r^2}\right )^{1/2}.
\ee
We define the truncated Arvis potential by expanding the potential in a 
power series and retaining the first three terms. 

The L\"uscher term $c(r)$ is determined locally at different values of $r$'s by
\be
c(r)= \frac{r^3}{2}\frac{\partial^2 V(r)}{\partial r^2}
\equiv \frac{r^3}{2}(V(r+1)-2 V(r)+V(r-1)).
\ee
The theoretical predictions in the continuum limit are 
\bea
c(r)&=& -\pi/24 \hspace*{4cm}{\rm free\; bosonic\; string} \\
 &=&-\frac{\pi}{24}\left(1+\frac{\pi}{8\sigma r^2}\right ) \hspace*{2cm}
{\rm truncated\; Arvis} \\
 &=& -\frac{\pi}{24}\left(1-\frac{\pi}{12\sigma r^2}\right)^{-3/2}. 
\hspace*{1.2cm}{\rm Arvis}
\eea

To study the continuum limit of the behaviour of $c(r)$ we need to look at the 
variation of $c(r)$ at a fixed physical distance for various values of the bare 
coupling. Since 
the lattice spacing varies with the coupling constant, to compare the data
at different values of $\beta$, we need to introduce a common scale. To that 
end let us define the Sommer scale $r_0$ by $r_0^2 F(r_0)=1.65.$ For each value
of $\beta$, the Sommer scale has been estimated and is shown in table \ref{tab0}.
We also see from the same table that $\sigma r_0^2$ is constant $(\simeq 1.23)$ to a very good 
approximation.

\begin{figure}[h]
\mbox{\epsfig{file=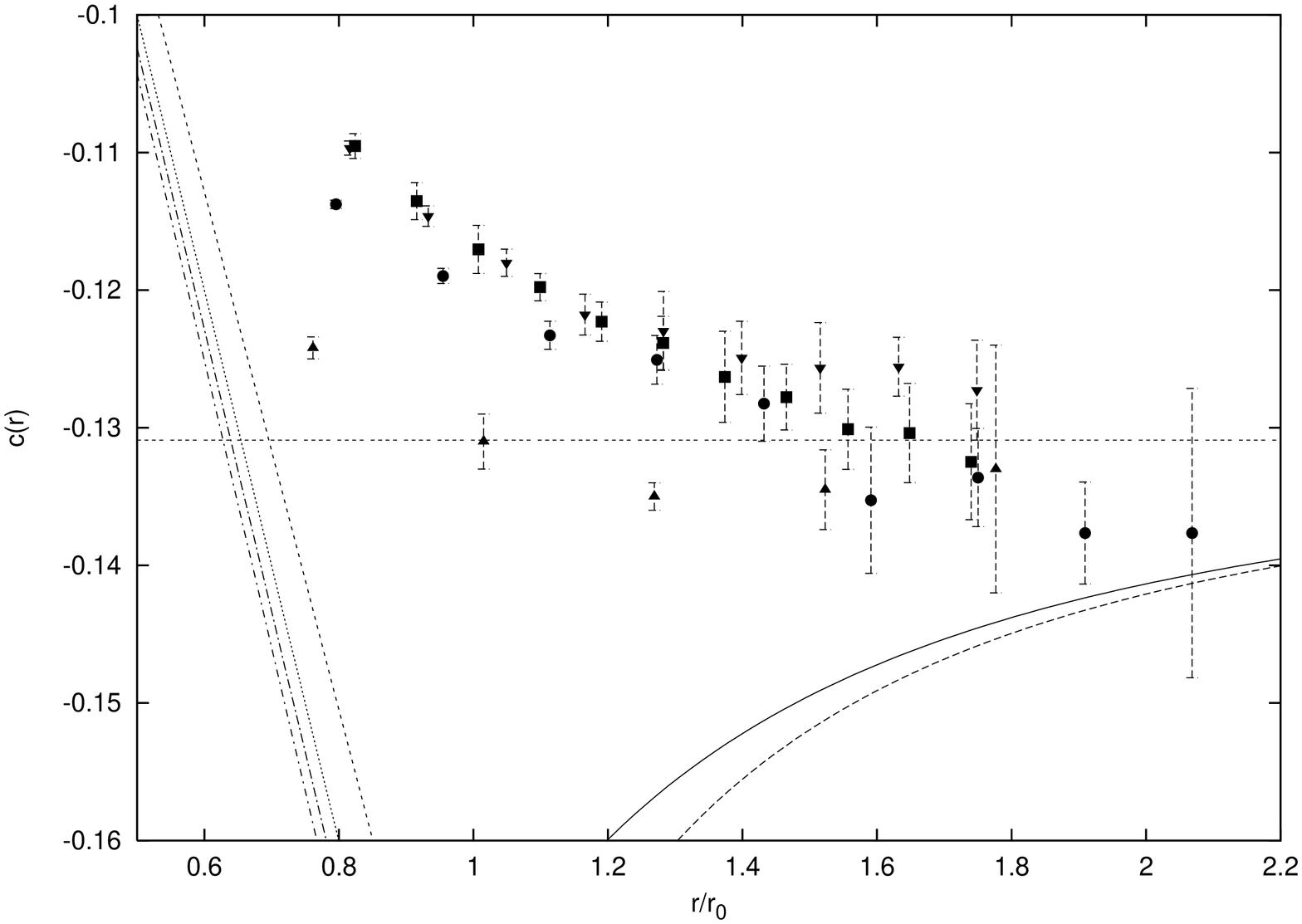,width=10truecm,angle=-90}}
\caption{$c(r)$ at various values of $\beta$.
$\blacktriangle$, $\bullet$, $\blacktriangledown$, $\blacksquare$ correspond 
to $\beta=$ 5.0, 7.5, 10.0 and 12.5 respectively. The 4 lines
denoted by $-\:-$, $\cdots$, $-\!\cdot\!-$ and $-\:\cdot\:-$ are
the corresponding 2-loop perturbation theory curves.
$----$ is $c(r)$ from free bosonic string
while the continuous line and the dashed curve are the values predicted by the
truncated Arvis and the Arvis potentials.}\label{cr.fig}
\end{figure}

We now plot all the data together in figure \ref{cr.fig} using $r/r_0$ as the 
x-axis. The data for $\beta=5$ has been taken from \cite{P1}.
On this figure we also plot the prediction of $c(r)$ from 2-loop 
perturbation theory given by \cite{York} 
\be
c(r/r_0)=-\frac{3(r/r_0)r_0}{4\pi\beta}.
\ee
Our data approaches the theoretical values from above. For $\beta=5$
 and $7.5$ the data points seem to cross the free string value and come closer
to the predictions from the Arvis potentials. For the other two $\beta$ values
the explored regions of $r$ unfortunately turned out to be too small to see if they 
too actually cross $-\pi/24$ and again reapproach that value from below. It 
would therefore be important to see what happens to $c(r)$ as one 
extends the range of $r$. 
    
It is also interesting to observe that the perturbative curve is closest to the 
lattice data for the largest lattice spacing. Thus it seems that the departure
from the perturbation theory sets in earlier as one goes towards the continuum
limit. Also we see that the data sets at $\beta=10.0$ and $\beta=12.5$ are 
degenerate within error bars. This of course makes it pointless to do a further
extrapolation to the continuum limit and we conclude that within our error 
bars the data at $\beta=12.5$ is identical with the continuum limit.

\section{Wilson loop}

Space-time Wilson loops of extent $(r,T)$ can be interpreted in the transfer 
matrix formalism, as the correlation of sources between points separated by
a distance $r$, propagating for time $T$.
While the Polyakov loop correlators do not give us the option of having 
sources and project strongly onto the ground state,
Wilson loops enable us to study the excited states by constructing
sources which project onto some state preferentially. 

In three dimensions the flux-tube has only one transverse direction to fluctuate 
and thus possesses only discrete symmetries according to the transformation 
properties under charge conjugation and parity. Classification of the low lying 
states according to these symmetries have been carried out in \cite{P1}, and using 
the sources described there, we study the ground state and the first excited state 
of the string 
spectrum and take a careful look at the energy difference between the two states, 
comparing them with the prediction from the free bosonic string and the Arvis 
potential. 

In our measurements, we compute correlation matrices $C(r,T)$ among different 
sources, each between points separated by a distance $r$,
and then diagonalize the correlation matrix by taking appropriate combinations 
of the matrix elements. Energies of the string states can then be obtained by 
fitting the eigenvalues $\lambda(r,T)$ of $C(r,T)$ to the form 
\be\label{naive}
\lambda(r,T) = \alpha e^{-E(r)T}
\ee
using different $T$ values. The energies obtained in this way are reported under
``no removal" in tables \ref{tab4} and \ref{tab5}.
This however is a naive estimate and in 
practice there are corrections due to higher excited states. That is because, 
while sources with different symmetry properties allow us to 
project onto different channels, there are still infinitely many states in each 
channel and they all contribute to the correlator in that channel. We are interested
only in the leading exponential in each of the channels, but even these are 
significantly contaminated by the higher states at the smaller values of $T$. 
To make a meaningful comparison with the prediction from the string picture,
we need to remove this contamination as much as possible.

\begin{figure}
\vspace*{-1cm}
\begin{center}
\mbox{\epsfig{file=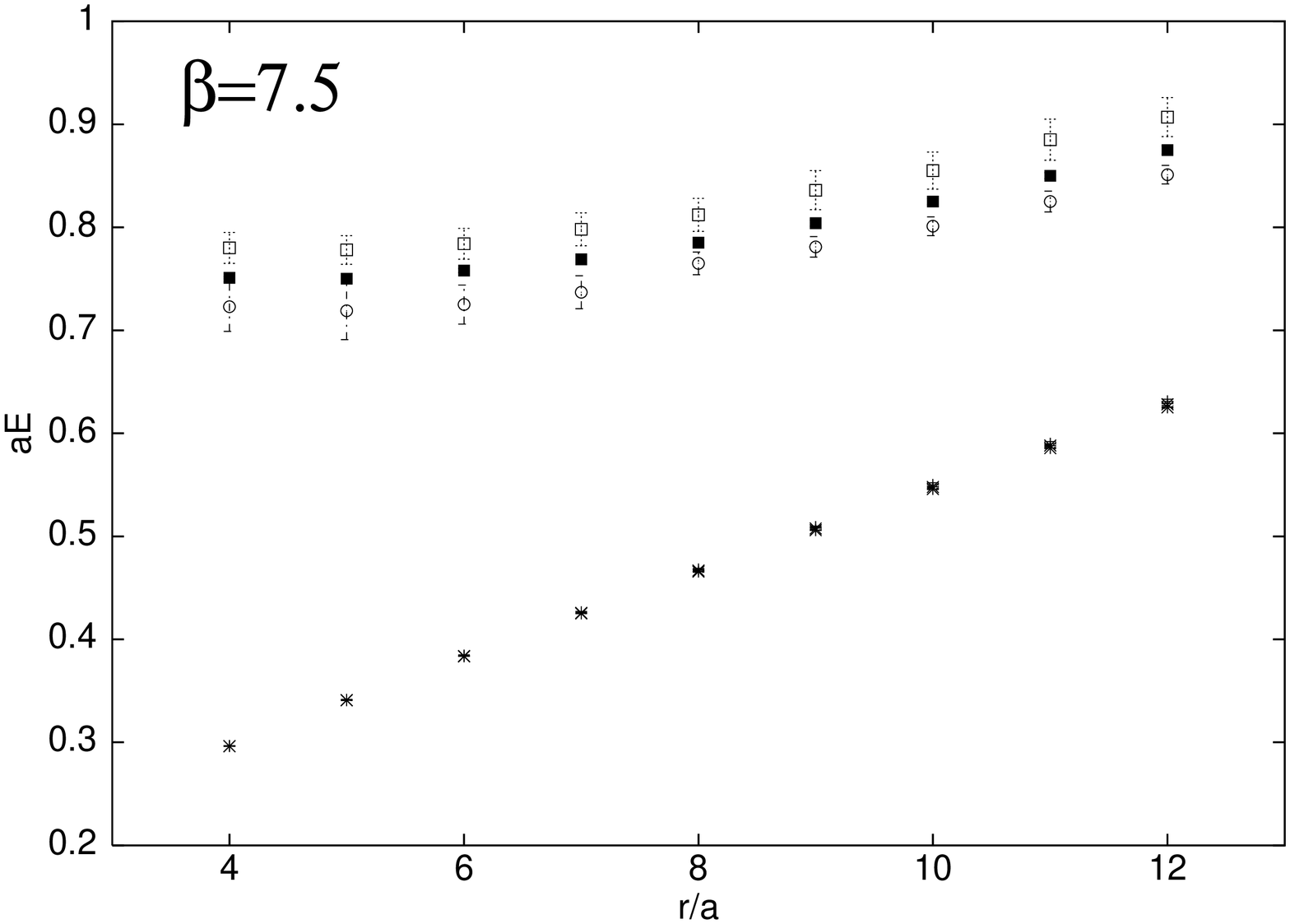,width=10truecm,angle=0}}
\mbox{\epsfig{file=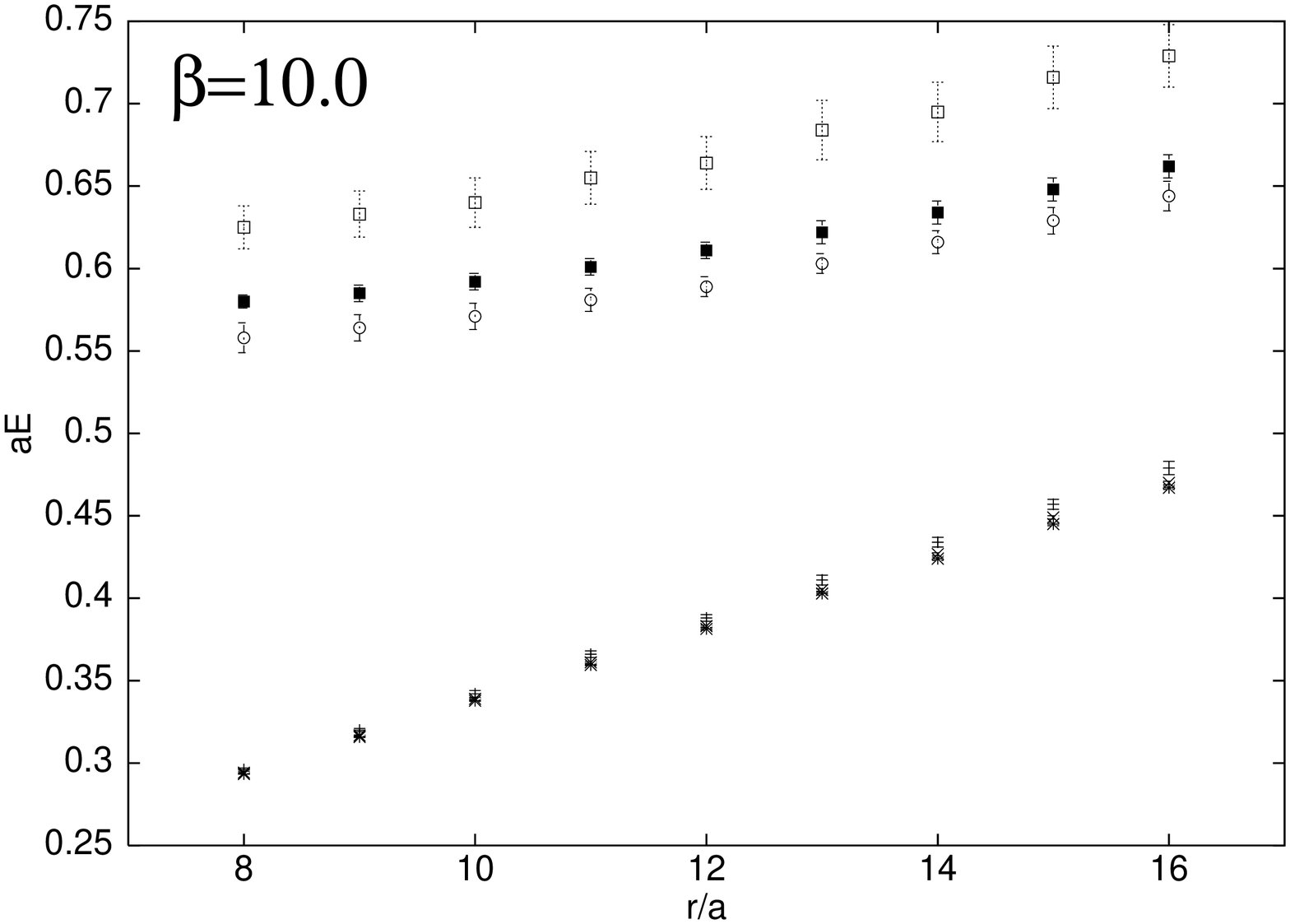,width=10truecm,angle=0}}
\mbox{\epsfig{file=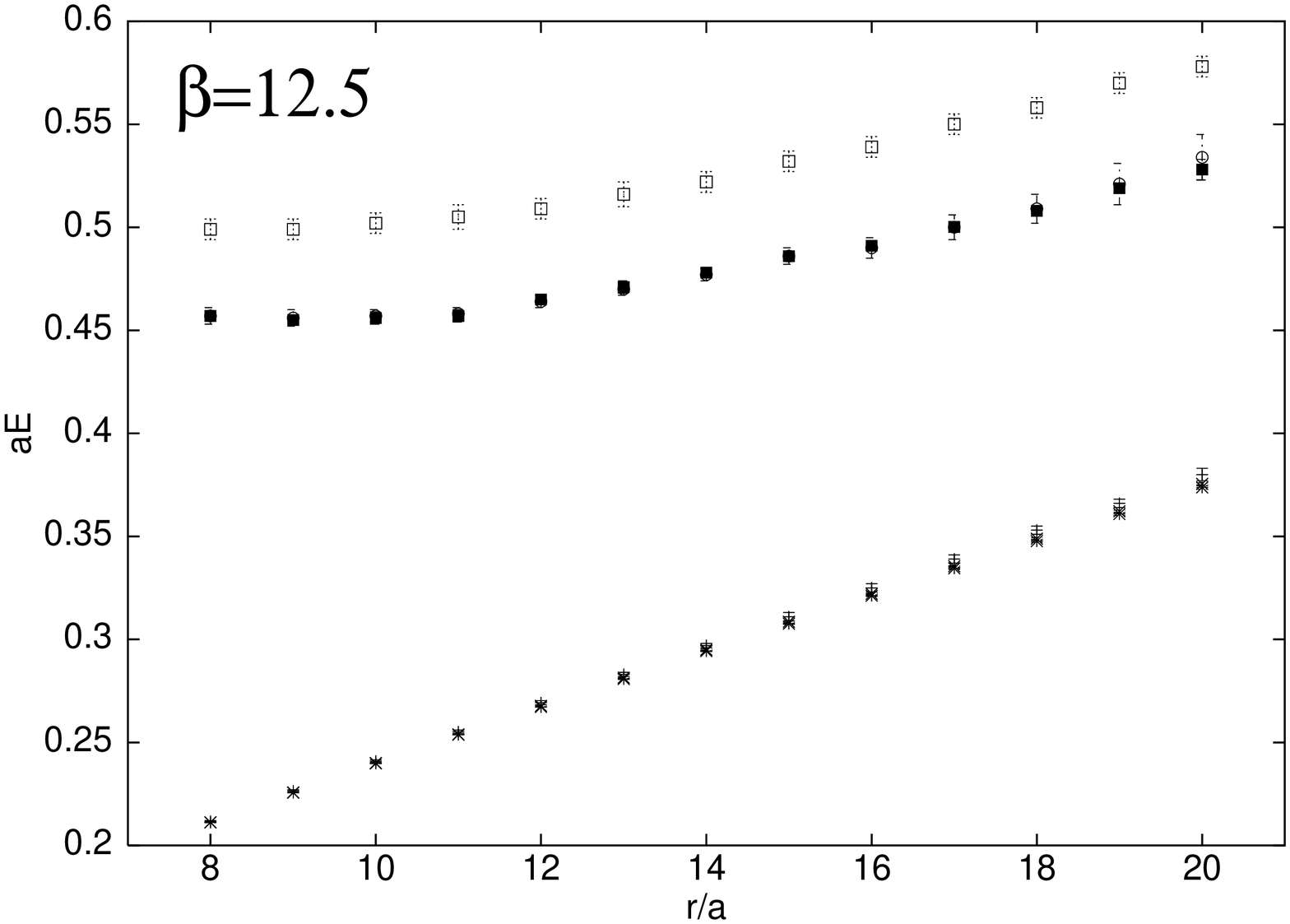,width=10truecm,angle=0}}
\end{center}
\caption{The first two energy states at various values of $\beta$.
The three sets of points are ``no removal", ``partial removal" and
``full removal" with ``no removal" being the highest and ``full removal"
being the lowest.}\label{en.fig}
\end{figure}

We do the removal in two ways. First we drop the smallest value of $T$ and do a
fit to the rest of the points. These values are reported under the head 
``partial removal". Second we assume that 
\be
\lambda(r,T) = a(r) e^{-E_0(r)T} + b(r) e^{-(E_0+\delta E_0)(r)T}
\ee
where $E_0+\delta E_0$ is the energy of the state above $E_0$ in the same
symmetry channel.
This allows us to  define an improved observable ${\bar E}$ by
\be
-\frac{1}{T_2-T_1}\log\frac{\lambda (T_2)}{\lambda (T_1)}
={\bar E} + \frac{1}{T_2-T_1}\left [ \frac{b(r)}{a(r)}e^{-\delta E_0 T_1}
(1-e^{-\delta E_0(T_2-T_1)})\right ].
\ee
We estimate that further corrections coming from 
higher states are smaller than our statistical errors. We denote the ${\bar E}$ 
values in the tables under ``full removal". We plot the first two states for the 
different $\beta$ values in figure \ref{en.fig}.

As seen from the figure, for the ground state the corrections
due to finite time extents are small and so all the three sets of points are on top
of each other. For the first excited state, the corrections are clearly visible.
 For $\beta=12.5$,
our data was not good enough to do the extrapolation for a ``full removal". Hence
it has only two sets of points for the excited state. In table \ref{tab5}, the points marked 
with a $\ast$ were also too unstable for extrapolation and the values reported there are the 
values extracted from the Wilson loops with the two largest time extents.

Let us now come to the energy difference between the ground state and the first
excited state. A free bosonic string description of the flux tube predicts this energy 
difference to be $\pi/r$ for a string formed between points separated by a distance
 $r$. The Arvis energy states are given by 
\be
E_n=\sigma r \sqrt{1+\frac{2\pi}{\sigma r^2}\left ( n-\frac{d-2}{24}\right )}
\ee
where $d$ is the number of space-time dimensions. It is interesting to observe 
that degeneracies of the Arvis states is the same as that of the free string \cite{LW2}
For us $d=3$ and therefore for $E_1-E_0$ 
we get 
\bea
E_1-E_0 &=& \frac{\pi}{r}\left( 1-\frac{11\pi}{24\sigma r^2}\right )\hspace*{2cm}{\rm 
(truncated \;Arvis)} \\
&=& \sigma r \left(\sqrt{1+\frac{23\pi}{12\sigma r^2}}
-\sqrt{1-\frac{\pi}{12\sigma r^2}}\right ) {\rm \;\;(Arvis)}
\eea

\begin{figure}
\vspace*{-1.5cm}
\begin{center}
\mbox{\epsfig{file=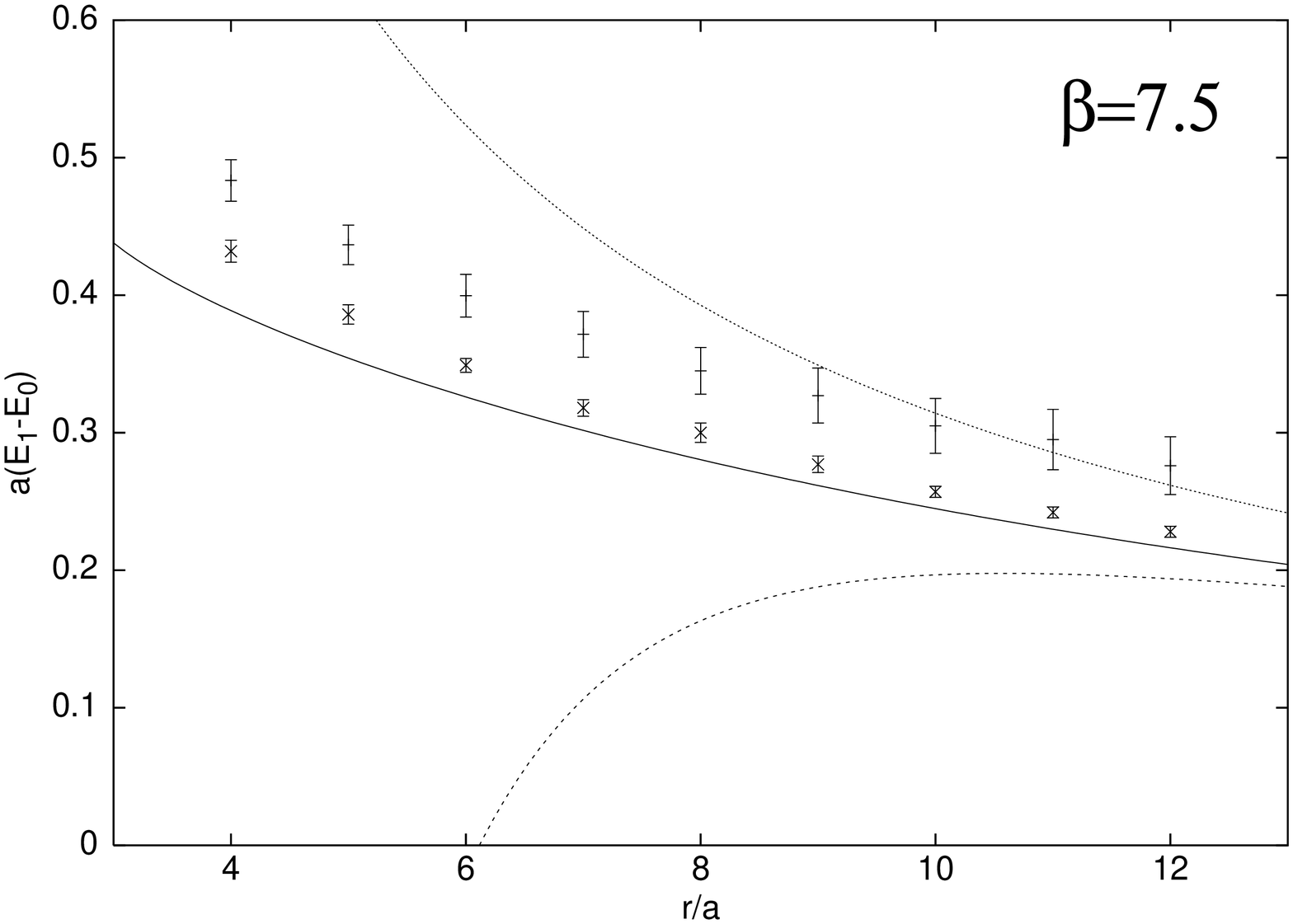,width=10truecm,angle=0}}
\mbox{\epsfig{file=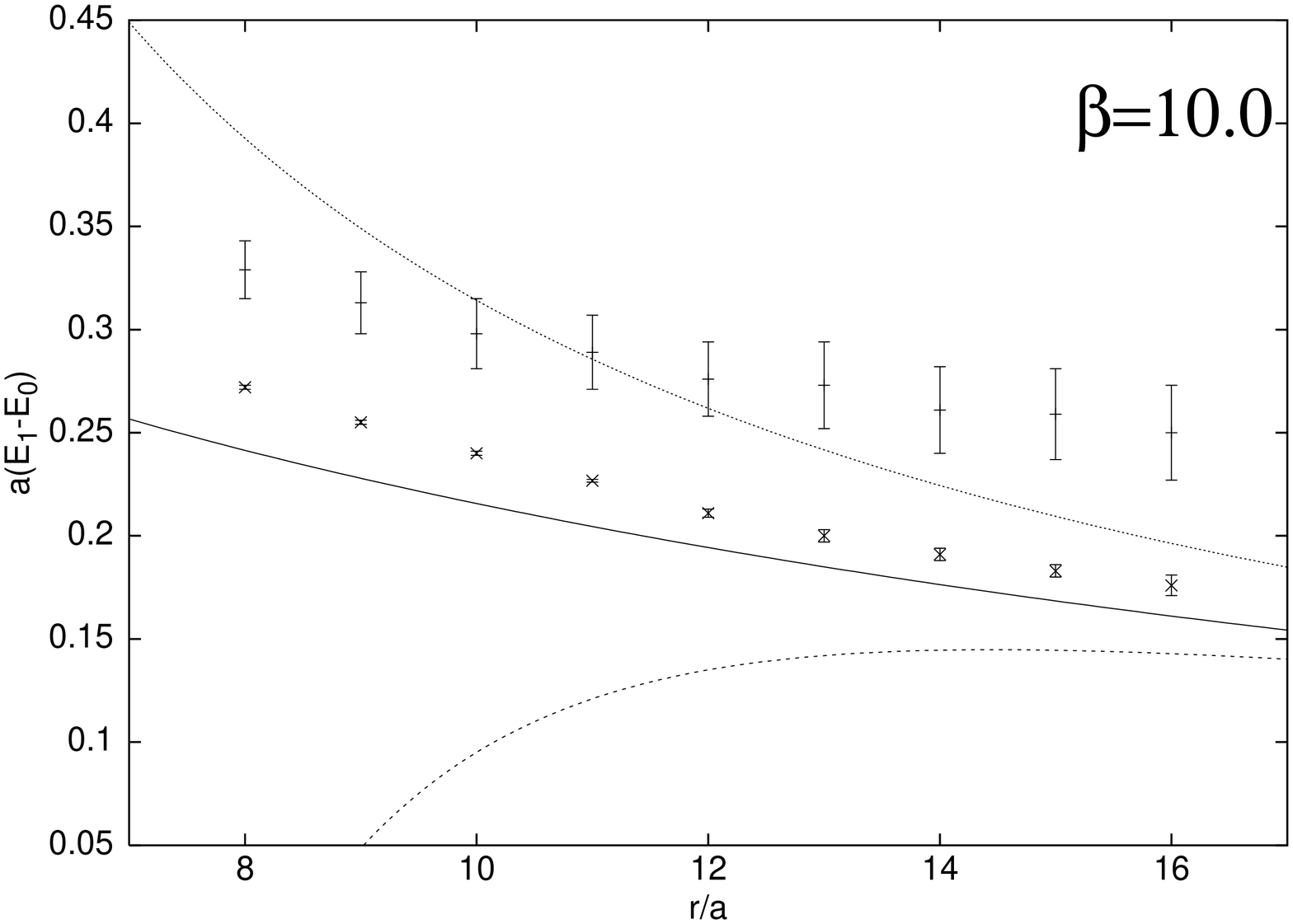,width=10truecm,angle=0}}
\mbox{\epsfig{file=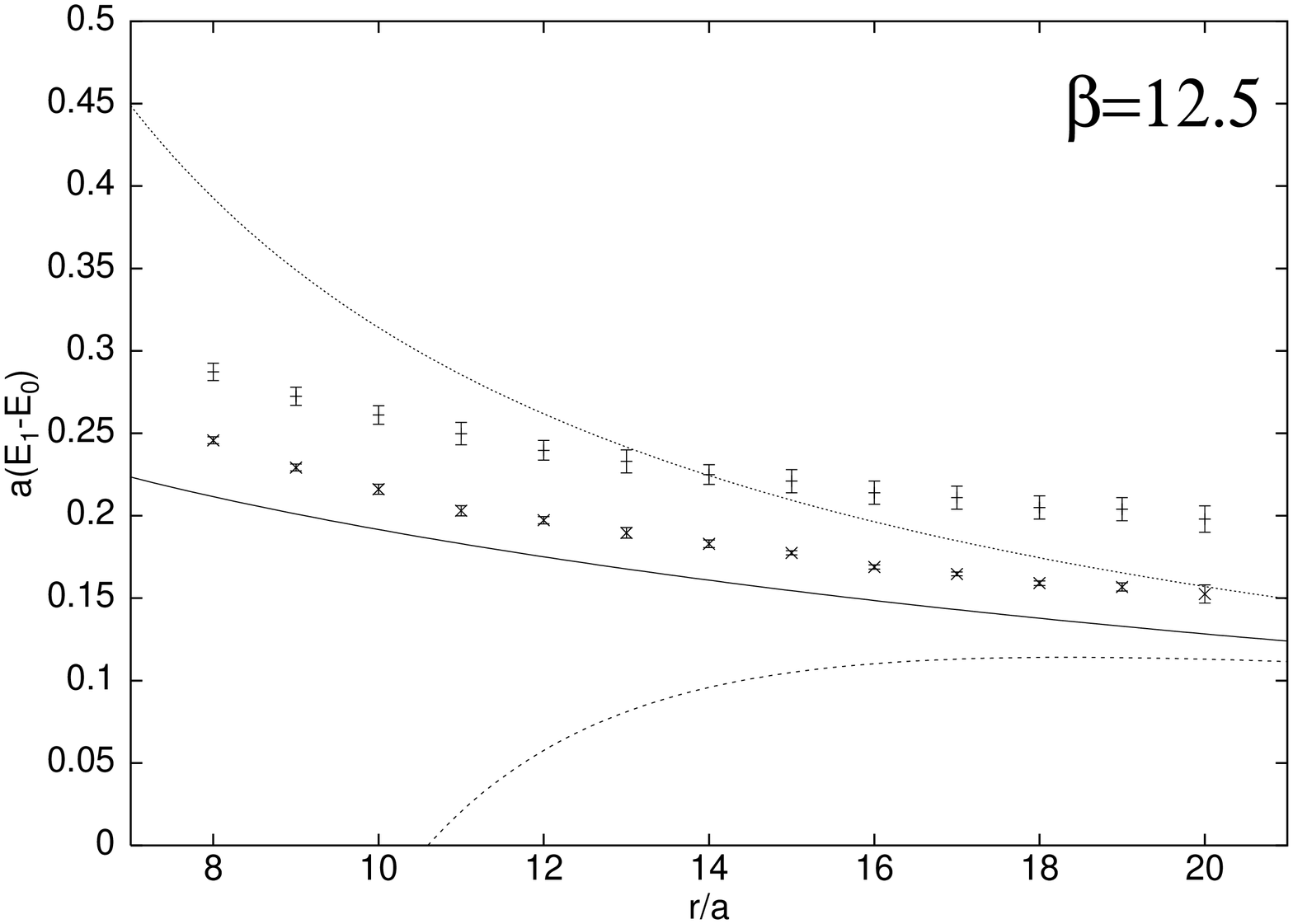,width=10truecm,angle=0}}
\end{center}
\caption{$E_1-E_0$ at various values of $\beta$. The two   sets of
points correspond to the energy differences computed from the sets
``no removal" (upper) and ``Correlated difference" (lower)
. The dotted curve is the free bosonic string prediction. The
dashed curve and the continuous curve correspond to the truncated and full Arvis
potentials respectively.}\label{ediff.fig}
\end{figure}

In figure \ref{ediff.fig}, we plot these energy differences along with the predictions
from the free bosonic string and Arvis potentials. 
The two sets of points are the differences computed in different ways. The upper set is just the naive energy difference obtained from the column ``no removal". The lower set is computed from
the ratios of the eigenvalues of the correlation matrices. To do this we need two correlators at 
two different times for each state. Suppressing the $r$ dependencies we denote these states by
\bea
c(t_1)&=&a\:e^{-E_0t_1}+b\:e^{-(E_0+\delta E_0)t_1}\\
c(t_2)&=&a\:e^{-E_0t_2}+b\:e^{-(E_0+\delta E_0)t_2}\\
d(t_1)&=&a^{'}e^{-E_1t_1}+b^{'}e^{-(E_1+\delta E_1)t_1}\\
d(t_2)&=&a^{'}e^{-E_1t_2}+b^{'}e^{-(E_1+\delta E_1)t_2}
\eea

Then it is easy to see that to leading order

\be
\hspace*{-1cm}\frac{1}{\Delta t}\log\left[\frac{c(t_1)/c(t_2)}{d(t_1)/d(t_2)}\right] = (E_1-E_0)+
\frac{1}{\Delta t}\left [ 
\frac{b}{a}e^{-\delta E_0 t_1}-\frac{b^{'}}{a^{'}}e^{-\delta E_1 t_1}-\frac{b}{a}e^{-\delta E_0 t_2}+\frac{b^{'}}{a^{'}}e^{-\delta E_1 t_2} \right ]
\ee
where $\Delta t=t_1-t_2$.
Thus we see that the corrections are exponentially suppressed. Moreover to leading order $\delta E_0$ and $\delta E_1$ are identical and thus a further cancellation takes place between the coefficients.
Here we use the fact that since the different correlators at the same times were measured in the
same simulation we take the errors with the same sign on them.
 Taking the two largest time extents we get values for $E_1-E_0$ consistent with what we would obtain from the columns ``full removal", but with lower errors. We will call the energy differences 
obtained in this way ``Correlated energy difference". 
In table \ref{tabediff} we tabulate the ``Correlated energy difference" for $E_1-E_0$. These
values are in agreement with values obtained elsewhere using 
asymmetric lattices \cite{Kuti2}. 

\begin{table}[htb]\caption{Correlated energy difference for $E_1-E_0$}\label{tabediff}
\begin{center}
\begin{tabular}{rclrcl}
\hline
\multicolumn{3}{c}{Lattice {\bf A}}&\multicolumn{3}{c}{Lattice {\bf B}} \\
\hline
$r$ & $E_1-E_0$ & error & $r$ & $E_1-E_0$ & error \\
\hline
4 & 0.432 & (8) & 8 & 0.272 & (1) \\
5 & 0.386 & (7) & 9 & 0.255 & (1) \\
6 & 0.349 & (5) & 10 & 0.240 & (1) \\
7 & 0.318 & (6) & 11 & 0.2267 & (7) \\
8 & 0.300 & (7) & 12 & 0.211 & (2) \\
9 & 0.277 & (6) & 13 & 0.200 & (3) \\
10 & 0.257 & (4) & 14 & 0.191 & (3) \\
11 & 0.242 & (4) & 15 & 0.183 & (3) \\
12 & 0.228 & (4) & 16 & 0.176 & (5) \\
\hline 
\end{tabular}
\end{center}
\end{table}

As can be seen from figure \ref{ediff.fig}, the variation of the energy difference with $r$
 is much better described by the Arvis potential than the free bosonic string for 
smaller values of $r$.  
In the range of $r$ we have looked at the systematic
corrections are very important. It is only the data set corrected for higher
energy states that stay close to the Arvis potential and keep approaching the free
string prediction as one goes to larger values of $r$. Without taking these 
corrections into account, one can draw wrong conclusions about the validity of the
string picture as our naive data actually crosses the free string value and the 
difference seems to increase with $r$ instead of decreasing. We expect as one
goes to larger values of $r$ even more careful treatment of the higher energy
corrections are going to be necessary.

\begin{figure}[htb]
\centerline{\mbox{\epsfig{file=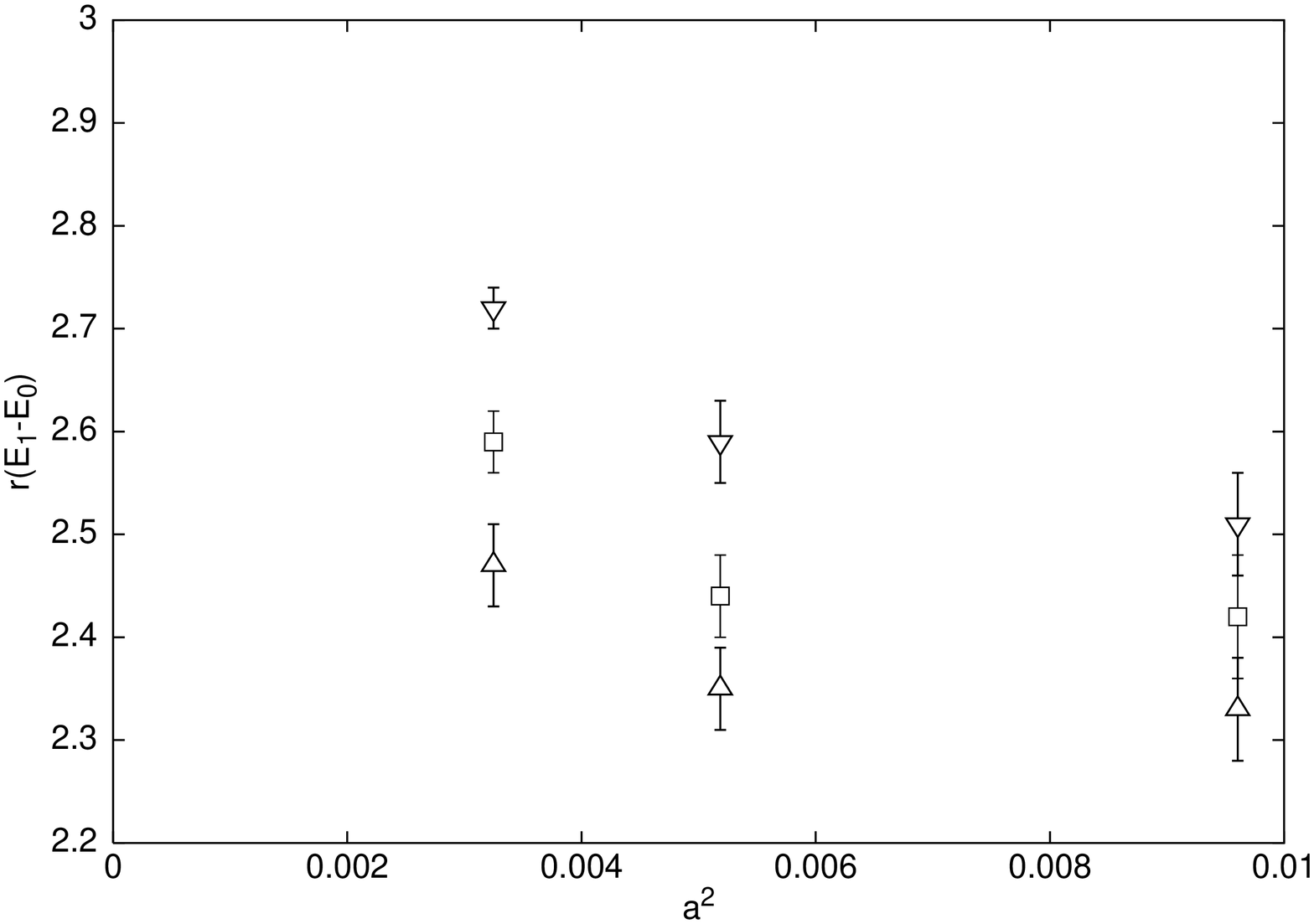,width=7truecm,angle=-90}}}
\caption{$r(E_1-E_0)$ against $a^2$. $\triangle$ , $\Box$ and $\triangledown$ correspond to
$r/r_0$=1.2, 1.3 and 1.5 respectively.}\label{edextra.fig}
\end{figure}

\begin{figure}[htb]
\centerline{\mbox{\epsfig{file=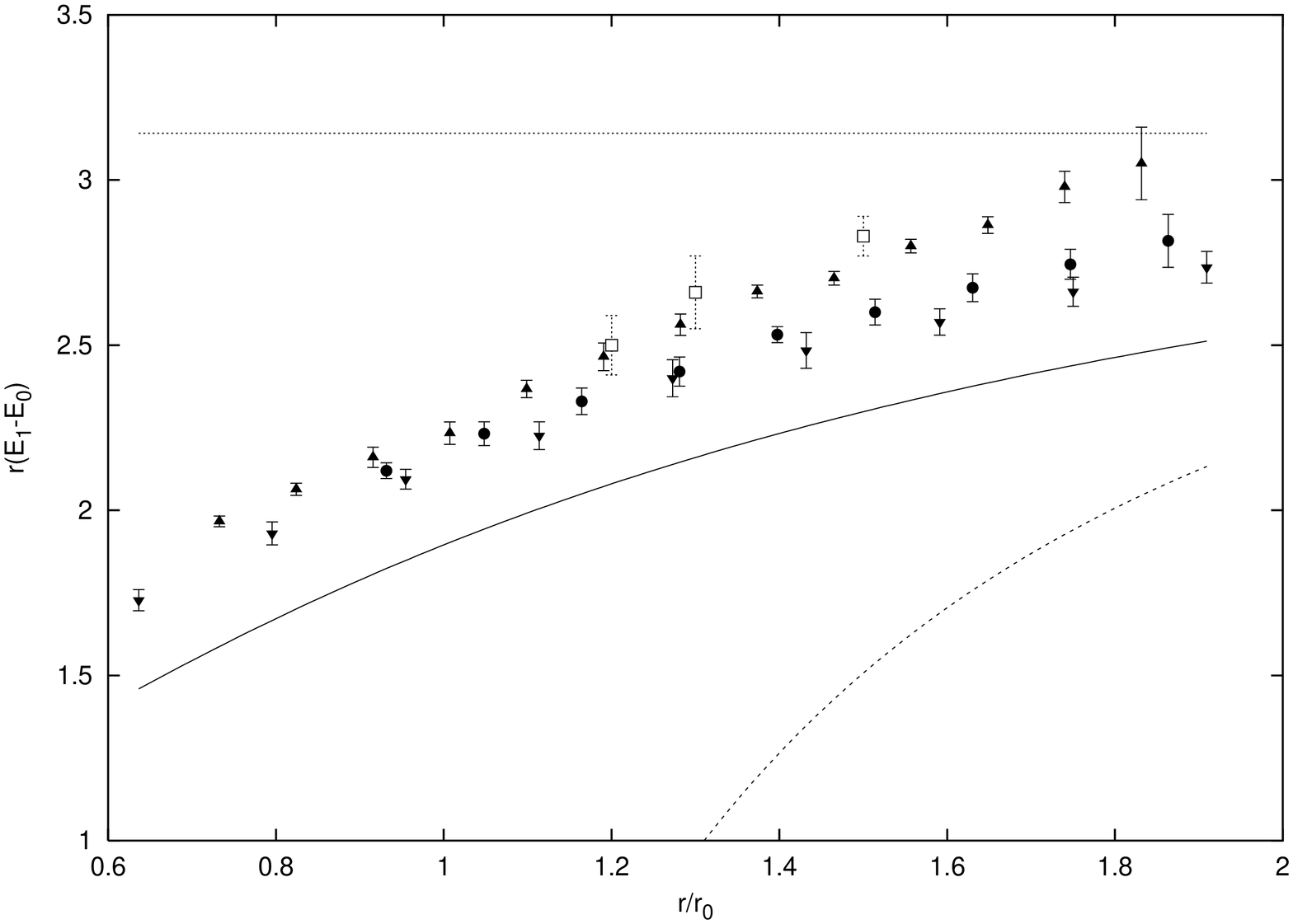,width=9truecm,angle=-90}}}
\caption{ $r(E_1-E_0)$ against $r/r_0$.
$\blacktriangledown$, $\bullet$ and $\blacktriangle$ correspond to $\beta$=7.5, 10.0 and 12.5 
respectively. The continuous curve is the prediction from the Arvis potential and the 
dashed curve the truncated Arvis one. The dotted line is the free string prediction.
The $\square$'s are the continuum limit values at $r/r_0=1.2,1.3$ and $1.5$ respectively.
}\label{edcont.fig}
\end{figure}

To have an idea of the magnitude of the lattice effects, we compare $r(E_1-E_0)$ for
the three different lattice spacings at $1.2r_0$, $1.3r_0$, and $1.5r_0$. The energy
 differences at these distances are obtained by interpolation. We plot these differences
against $a^2$ in figure \ref{edextra.fig}. If we assume that the continuum limit is
approached as $a^2$, then we can fit a straight line through these points to obtain
 the continuum limit values of the energy difference. We tabulate these values in
table \ref{conlim}.

In figure \ref{edcont.fig}, we plot $r(E_1-E_0)$ against $r/r_0$ for all the three values
of $\beta$ that we have investigated along with the continuum limit values at $r/r_0$=
1.2, 1.3 and 1.5. 
We see that the entire data set
lies between the free string value and the values from the Arvis potential. Lattice
effects are present but do not seem to be too large. As in the case of $c(r)$, the continuum 
limit values are not too far from the lattice at $\beta$=12.5. The direction of the shift with smaller lattice spacing 
seems to be away from the Arvis and towards the free string description. 
Also as we have mentioned before, for $\beta=12.5$ we only have a partial removal of the 
higher states for the first excited state and hence the estimated energy differences 
are higher than the true ones. Therefore we our values indicate only an upper bound for the 
continuum limit rather than actual values. 

\begin{table}\caption{Continuum limit extrapolation of $r(E_1-E_0)$}\label{conlim}
\begin{center}
\vspace*{-5mm}\begin{tabular}{c|c|c|c|c}
\hline
& $a^2=0.009666$ & $a^2=0.005162$ & $a^2=0.003206$ & $a^2=0$ \\
\hline
1.2$r_0$ & 2.33 (5) & 2.35 (4) & 2.47 (4) & 2.50 (9) \\
1.3$r_0$ & 2.42 (6) & 2.44 (4) & 2.59 (3) & 2.66 (11) \\
1.5$r_0$ & 2.51 (5)& 2.59 (4)& 2.72 (2) & 2.83 (6) \\
\hline
\end{tabular}
\end{center}
\end{table}

\begin{figure}[htb]
\centerline{\mbox{\epsfig{file=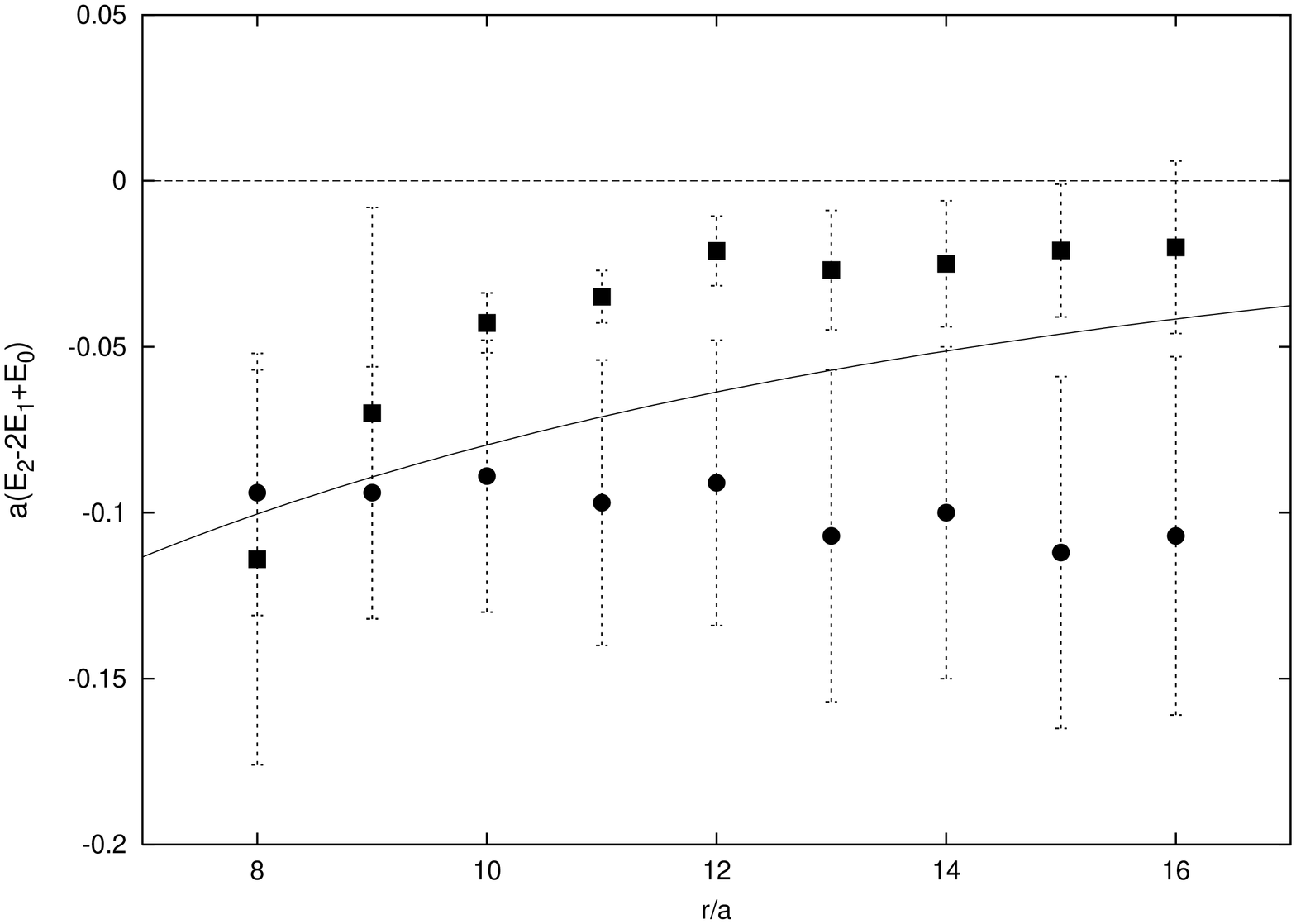,width=8truecm,angle=-90}}}
\caption{$(E_2-E_0)-2(E_1-E_0)$ against $r/a$ at $\beta=10.0$.
The free bosonic string predicts the value zero while the other curve
 is the Arvis potential. $\bullet$, and $\blacksquare$ 
correspond to the sets ``no removal" and ``full removal" 
respectively. 
}\label{2ediff.fig}
\end{figure}

\section{Conclusions}
In this paper we have looked at the continuum limit of the L\"uscher term and the energy
 difference between the ground state and the first excited state.
Our study indicates that for the L\"uscher term, the asymptotic value is approached in a
non-monotonic way. 
 This non-monotonicity is seen on our coarser lattices.
On our finer lattices, the range of $r$ looked at and big error bars prevent drawing any
definite conclusion but similar tendencies are present even on our finest lattice.

The energy difference at finite lattice spacing seems to be well described by the Arvis 
potential. However the continuum extrapolation pushes it away towards the free string
 description. It is very important to take into account the corrections due to the higher
energy states. Otherwise one has a large systematic error and the resulting energy 
difference seems to diverge from either description rather than tend to it with increasing 
$r$.
Another important observable is $(E_2-E_0)-2(E_1-E_0)$. We plot this 
quantity at $\beta=10$ in figure \ref{2ediff.fig}.
In spite of large error bars, the behaviour of the ``full removal" data set is consistent
with the prediction from the Arvis potential. The ``no removal" set on the other hand 
has a qualitatively 
different behaviour.

We are finally in a position to start distinguishing between different string 
models as the subleading effects are visible and it will be really interesting
to extend the study to larger $r$ on the finer lattices.  

\section{Acknowledgments}
During the course of this work, the author was supported by Fonds zur F\"orderung
der Wissenschaftlichen Forschung in \"Osterreich, under the Lise Meitner project 
M767-N08. 
The author would also like to thank Peter Weisz for several useful discussions and 
for suggesting the comparison with the Arvis potential. The simulations were partly 
carried out on the Linux Cluster at The Institute of Mathematical Sciences, Chennai, 
India. The author is indebted to the institute for this facility.

\newpage
\addtolength{\headsep}{-1.5cm}
\addtolength{\footskip}{3.5cm}

\begin{table}\caption{Polyakov loop data}\label{tab3}
\begin{center}
\begin{tabular}{llll}
\hline
\multicolumn{4}{l}{$lattice$ {\bf A}} \\
\hline
$r$ & $V(r)$ & $F(r)$ & $c(r)$ \\
4 & 0.29620 (4) & $\;\;\;\;-$ & $\;\;\;\;-$ \\
5 & 0.34065 (7) & 0.04353 (2) & $-$0.1138 (3) \\
6 & 0.38327 (9) & 0.04207 (3) & $-$0.1190 (5) \\
7 & 0.4248 (1) & 0.04116 (3) & $-$0.123 (1) \\
8 & 0.4656 (2) & 0.04069 (6) & $-$0.125 (2) \\
9 & 0.5067 (3) & 0.04027 (6) & $-$0.128 (3) \\
10 & 0.5468 (3) & 0.03996 (7) & $-$0.135 (5) \\
11 & 0.58601 (7) & 0.03965 (2) & $-$0.134 (4) \\
12 & 0.62556 (8) & 0.03947 (2) & $-$0.138 (4) \\
13 & 0.6649 (1) & 0.03933 (2) & $-$0.138 (10) \\
14 & 0.7042 (1) & $\;\;\;\;-$ & $\;\;\;\;-$
\vspace*{5mm}
\end{tabular}

\begin{tabular}{llll}
\hline
\multicolumn{4}{l}{$lattice$ {\bf B}} \\
\hline
$r$ & $V(r)$ & $F(r)$ & $c(r)$ \\
6 & 0.24727 (8) & $\;\;\;\;-$ & $\;\;\;\;-$ \\
7 & 0.2707 (1) & 0.02315 (3) & $-$0.1097 (5) \\
8 & 0.2935 (1) & 0.02261 (3) & $-$0.1146 (7) \\
9 & 0.3158 (2) & 0.02223 (4) & $-$0.118 (1) \\
10 & 0.3378 (2) & 0.02195 (4) & $-$0.122 (1) \\
11 & 0.3595 (2) & 0.02174 (4) & $-$0.123 (3) \\
12 & 0.3811 (3) & 0.02151 (4) & $-$0.125 (3) \\
13 & 0.4027 (3) & 0.02138 (4) & $-$0.126 (3) \\
14 & 0.4240 (4) & 0.02128 (2) & $-$0.126 (2) \\
15 & 0.4452 (2) & 0.02120 (2) & $-$0.127 (4) \\
16 & 0.4664 (2) & $\;\;\;\;-$ & $\;\;\;\;-$
\vspace*{5mm}
\end{tabular}

\begin{tabular}{llll}
\hline
\multicolumn{4}{l}{$lattice$ {\bf C}} \\
\hline
$r$ & $V(r)$ & $F(r)$ & $c(r)$ \\
8 &  0.21098 (9) & $\;\;\;\;-$ & $\;\;\;\;-$ \\
9 &  0.2254 (1) & 0.01428 (2) & $-$0.1095 (9) \\
10 & 0.2395 (1) & 0.01401 (3) & $-$0.114 (1) \\
11 & 0.2536 (1) & 0.01381 (3) & $-$0.117 (2) \\
12 & 0.2673 (1) & 0.01368 (3) & $-$0.120 (1) \\
13 & 0.2809 (2) & 0.013555 (8) & $-$0.122 (1) \\
14 & 0.2944 (2) & 0.013454 (9) & $-$0.124 (2) \\
15 & 0.3079 (2) & 0.01337 (1) & $-$0.126 (3) \\
16 & 0.3212 (2) & 0.01332 (2) & $-$0.128 (2) \\
17 & 0.3346 (2) & 0.01326 (2) & $-$0.130 (3) \\
18 & 0.3479 (2) & 0.01322 (3) & $-$0.130 (4) \\
19 & 0.3611 (2) & 0.01318 (3) & $-$0.132 (4) \\
20 & 0.3742 (2) & $\;\;\;\;-$ & $\;\;\;\;-$
\end{tabular}
\end{center}
\end{table}

\begin{table}\caption{Ground state from Wilson loops for
lattices {\bf A, B} \& {\bf C}}\label{tab4}
\begin{center}
\begin{tabular}{l|lll|lll|ll}
\hline
{\bf A}& \multicolumn{3}{c|}{no removal} & \multicolumn{3}{c|}{partial removal}
&\multicolumn{2}{c}{full removal} \\
\hline
$r$ & E & $\Delta$E & $\chi^2/d.o.f$ & E & $\Delta$E & $\chi^2/d.o.f$ & E &
$\Delta$E \\
\hline
4 & 0.2966 & 0.0001 & 24.4 & 0.29640 & 0.00004 & 1.24 &  0.29631 & 0.00007 \\
5 & 0.3414 & 0.0003 & 75.3 & 0.34102 & 0.00009 & 4.25 &  0.3408  & 0.0001 \\
6 & 0.3844 & 0.0005 & 114  & 0.3838  & 0.0001  & 6.85 &  0.3835  & 0.0001 \\
7 & 0.4265 & 0.0007 & 202  & 0.4257  & 0.0002  & 12.3 &  0.4251  & 0.0002 \\
8 &   0.468  & 0.001  & 242 &  0.4668  & 0.0004 & 27.3 &  0.4657 & 0.0004 \\
9 &   0.509  & 0.001  & 329 &  0.5076  & 0.0006 & 37.3 &  0.5060 & 0.0004 \\
10 &  0.550  & 0.002  & 341 &  0.5479  & 0.0007 & 41.1 &  0.5460 & 0.0006 \\
11 &  0.590  & 0.002  & 400 &  0.5882  & 0.0009 & 50.2 &  0.5857 & 0.0007 \\
12 &  0.631  & 0.002  & 380 &  0.6282  & 0.0011 & 51.5 &  0.6251 & 0.0008
\vspace*{5mm}
\end{tabular}

\begin{tabular}{l|lll|lll|ll}
\hline
{\bf B} & \multicolumn{3}{c|}{no removal} & \multicolumn{3}{c|}{partial removal}
&\multicolumn{2}{c}{full removal} \\
\hline
$r$ & E & $\Delta$E & $\chi^2/d.o.f$ & E & $\Delta$E & $\chi^2/d.o.f$ & E &
$\Delta$E \\
\hline
8  &  0.296 & 0.001 & 542 & 0.2941 & 0.0002 & 8.5  & 0.2934 & 0.0002 \\
9  &  0.320 & 0.001 & 708 & 0.3167 & 0.0003 & 11.8 & 0.3158 & 0.0003 \\
10 &  0.342 & 0.002 & 764 & 0.3390 & 0.0004 & 13.9 & 0.3378 & 0.0004 \\
11 &  0.366 & 0.002 & 878 & 0.3611 & 0.0005 & 17.0 & 0.3595 & 0.0005 \\
12 &  0.388 & 0.002 & 883 & 0.3830 & 0.0007 & 18.4 & 0.3811 & 0.0005 \\
13 &  0.411 & 0.003 & 959 & 0.4050 & 0.0007 & 18.0 & 0.4029 & 0.0009 \\
14 &  0.434 & 0.003 & 927 & 0.4268 & 0.0008 & 18.8 & 0.424  & 0.001 \\
15 &  0.457 & 0.003 & 940 & 0.449  & 0.001  & 20.2 & 0.445  & 0.001 \\
16 &  0.479 & 0.004 & 879 & 0.470  & 0.001  & 20.0 & 0.467  & 0.001
\vspace*{5mm}
\end{tabular}

\begin{tabular}{l|lll|lll|ll}
\hline
{\bf C} & \multicolumn{3}{c|}{no removal} & \multicolumn{3}{c|}{partial removal}
&\multicolumn{2}{c}{full removal} \\
\hline
$r$ & E & $\Delta$E & $\chi^2/d.o.f$ & E & $\Delta$E & $\chi^2/d.o.f$ & E &
$\Delta$E \\
\hline
8   &   0.2117 & 0.0003 & 220 &   0.21122 & 0.00003 & 0.93 &   0.21115 & 0.00007 \\
9   &   0.2265 & 0.0005 & 317 &   0.22573 & 0.00005 & 1.51 &   0.22562 & 0.00009 \\
10  &   0.2409 & 0.0006 & 387 &   0.23994 & 0.00006 & 2.04 &   0.2398  & 0.0001 \\
11  &   0.2552 & 0.0008 & 480 &   0.25395 & 0.00009 & 2.77 &   0.2537  & 0.0001 \\
12  &   0.2693 & 0.0010 & 631 &   0.2677  & 0.0002  & 8.28  &  0.2672  & 0.0001 \\
13  &   0.283  & 0.001  & 715 &   0.2814  & 0.0002  & 9.75  &  0.2808  & 0.0002 \\
14  &   0.297  & 0.001  & 751 &   0.2950  & 0.0003  & 10.9  &  0.2943  & 0.0002 \\
15  &   0.311  & 0.002  & 804 &   0.3085  & 0.0003  & 12.3  &  0.3076  & 0.0003 \\
16  &   0.325  & 0.002  & 808 &   0.3221  & 0.0003  & 10.2  &  0.3212  & 0.0004 \\
17  &   0.339  & 0.002  & 838 &   0.3355  & 0.0003  & 11.0  &  0.3345  & 0.0005 \\
18  &   0.353  & 0.002  & 827 &   0.3489  & 0.0004  & 11.5  &  0.3477  & 0.0006 \\
19  &   0.366  & 0.002  & 832 &   0.3622  & 0.0005  & 12.3  &  0.3608  & 0.0007 \\
20  &   0.380  & 0.003  & 805 &   0.3755  & 0.0005  & 12.9  &  0.3738  & 0.0008
\end{tabular}
\end{center}
\end{table}

\begin{table}\caption{First excited state from Wilson
loops for lattices {\bf A, B} \& {\bf C}}\label{tab5}
\begin{center}
\begin{tabular}{l|lll|lll|ll}
\hline
{\bf A} & \multicolumn{3}{c|}{no removal} & \multicolumn{3}{c|}{partial removal}
&\multicolumn{2}{c}{full removal} \\
\hline
$r$ & E & $\Delta$E & $\chi^2/d.o.f$ & E & $\Delta$E & $\chi^2/d.o.f$ & E &
$\Delta$E \\
\hline
4  &   0.780 & 0.015 & 552 &   0.751 & 0.001 & 1.38 &   0.723 & 0.024 \\
5  &   0.778 & 0.014 & 446 &   0.750 & 0.002 & 2.15 &   0.719 & 0.028 \\
6  &   0.784 & 0.015 & 498 &   0.758 & 0.002 & 4.78 &   0.725 & 0.019 \\
7  &   0.798 & 0.016 & 498 &   0.769 & 0.003 & 5.35 &   0.737 & 0.016 \\
8  &   0.813 & 0.016 & 516 &   0.785 & 0.002 & 4.93  &  0.763 & 0.009 \\
9  &   0.836 & 0.019 & 545 &   0.804 & 0.003 & 5.23  &  0.781 & 0.010 \\
10 &   0.855 & 0.018 & 479 &   0.825 & 0.003 & 7.29  &  0.801 & 0.009 \\
11 &   0.885 & 0.020 & 447 &   0.850 & 0.003 & 6.33  &  0.825 & 0.010 \\
12 &   0.907 & 0.019 & 341 &   0.875 & 0.004 & 5.81  &  0.851 & 0.009
\vspace*{5mm}
\end{tabular}

\begin{tabular}{l|lll|lll|ll}
\hline
{\bf B} & \multicolumn{3}{c|}{no removal} & \multicolumn{3}{c|}{partial removal}
&\multicolumn{2}{c}{full removal} \\
\hline
$r$ & E & $\Delta$E & $\chi^2/d.o.f$ & E & $\Delta$E & $\chi^2/d.o.f$ & E &
$\Delta$E \\
\hline
8  &   0.625 & 0.013 & 938 & 0.580 & 0.004 & 20.6 & 0.558 & 0.009 \\
9  &   0.633 & 0.014 & 885 & 0.585 & 0.005 & 16.4 & 0.564 & 0.008 \\
10 &   0.640 & 0.015 & 833 & 0.592 & 0.005 & 16.8 & 0.571 & 0.008 \\
11 &   0.655 & 0.016 & 781 & 0.601 & 0.005 & 13.4 & 0.581 & 0.007 \\
12 &   0.664 & 0.016 & 690 & 0.611 & 0.005 & 13.5 & 0.589 & 0.006 \\
13 &   0.684 & 0.018 & 696 & 0.622 & 0.007 & 19.3 & 0.603 & 0.006(*) \\
14 &   0.695 & 0.018 & 587 & 0.634 & 0.007 & 15.6 & 0.616 & 0.007(*) \\
15 &   0.716 & 0.019 & 499 & 0.648 & 0.007 & 12.6 & 0.629 & 0.008(*) \\
16 &   0.729 & 0.019 & 328 & 0.662 & 0.007 & 9.1  & 0.644 & 0.009(*)
\vspace*{5mm}
\end{tabular}

\begin{tabular}{l|lll|lll}
\hline
{\bf C} & \multicolumn{3}{c|}{no removal} & \multicolumn{3}{c}{partial removal} \\
\hline
$r$ & E & $\Delta$E & $\chi^2/d.o.f$ & E & $\Delta$E & $\chi^2/d.o.f$ \\
\hline
8  &   0.499 & 0.005 & 58.8 &   0.457 & 0.002 & 0.45  \\
9  &   0.499 & 0.005 & 57.5 &   0.455 & 0.002 & 0.38  \\
10 &   0.502 & 0.005 & 59.6 &   0.456 & 0.003 & 0.68  \\
11 &   0.505 & 0.006 & 56.9 &   0.457 & 0.003 & 0.67  \\
12  &  0.509 & 0.005 & 48.1 &   0.465 & 0.002 & 0.19  \\
13  &  0.516 & 0.006 & 39.1 &   0.471 & 0.003 & 0.34  \\
14  &  0.522 & 0.005 & 31.7 &   0.478 & 0.002 & 0.18  \\
15  &  0.532 & 0.005 & 23.5 &   0.486 & 0.001 & 0.06  \\
16 &   0.539 & 0.005 & 23.0  &  0.491 & 0.001 & 0.03  \\
17 &   0.550 & 0.005 & 17.4  &  0.5002& 0.0009& 0.02  \\
18 &   0.558 & 0.005 & 13.8  &  0.508 & 0.001 & 0.04  \\
19 &   0.570 & 0.005 & 10.1  &  0.519 & 0.002 & 0.08  \\
20 &   0.578 & 0.005 & 8.2   & 0.528  & 0.005 & 0.34
\end{tabular}
\end{center}
\end{table}

\end{document}